\documentclass[]{spie}  

\usepackage{amsmath,amsfonts,amssymb}
\usepackage{graphicx}
\usepackage[colorlinks=true, allcolors=blue]{hyperref}

\PassOptionsToPackage{warn}{textcomp}
\usepackage[utf8]{inputenc}
\ifdefined\DeclareUnicodeCharacter
  \ifdefined\DeclareUnicodeCharacterAsOptional
    \def\sphinxDUC#1{\DeclareUnicodeCharacter{"#1}}
  \else
    \let\sphinxDUC\DeclareUnicodeCharacter
  \fi
  \sphinxDUC{00A0}{\nobreakspace}
  \sphinxDUC{2500}{\sphinxunichar{2500}}
  \sphinxDUC{2502}{\sphinxunichar{2502}}
  \sphinxDUC{2514}{\sphinxunichar{2514}}
  \sphinxDUC{251C}{\sphinxunichar{251C}}
  \sphinxDUC{2572}{\textbackslash}
\fi
\usepackage{cmap}
\usepackage[T1]{fontenc}
\usepackage{amsmath,amssymb,amstext}
\usepackage{babel}

\usepackage{times}
\expandafter\ifx\csname T@LGR\endcsname\relax
\else
  \substitutefont{LGR}{\rmdefault}{cmr}
  \substitutefont{LGR}{\sfdefault}{cmss}
  \substitutefont{LGR}{\ttdefault}{cmtt}
\fi
\expandafter\ifx\csname T@X2\endcsname\relax
  \expandafter\ifx\csname T@T2A\endcsname\relax
  \else
    \substitutefont{T2A}{\rmdefault}{cmr}
    \substitutefont{T2A}{\sfdefault}{cmss}
    \substitutefont{T2A}{\ttdefault}{cmtt}
  \fi
\else
  \substitutefont{X2}{\rmdefault}{cmr}
  \substitutefont{X2}{\sfdefault}{cmss}
  \substitutefont{X2}{\ttdefault}{cmtt}
\fi

\usepackage{sphinx}

\fvset{fontsize=\small}
\usepackage{geometry}

\usepackage{hyperref}
\usepackage{hypcap}
\addto\captionsenglish{}

\usepackage{sphinxmessages}

\date{}                           

\title{The adaptive optics simulation analysis tool(kit) (AOSAT)}

\author[a]{M. Feldt}
\author[a]{S. Hippler}
\author[a]{F. Cantalloube}
\author[a]{T. Bertram}
\author[b]{A. Obereder}
\author[a]{H. Steuer}
\author[c]{O. Absil}
\author[d]{M. Le Louarn}

\affil[a]{Max Planck Institute for Astronomy, Königstuhl 17, D-69117 Heidelberg, Germany}
\affil[b]{Johann Radon Institute, Altenberger Straße 69, A-4040 Linz, Austria }
\affil[c]{Planetary \& Stellar systems Imaging Laboratory, Université de Liège, Allée du 6 Août, 19C - Bât. B5c, B-4000 Liège 1, Belgium}
\affil[d]{European Southern Observatory, Karl-Schwarzschild-Str. 2, D-85748. Garching, Germany}

\authorinfo{Further author information: (Send correspondence to M. Feldt)\\M. Feldt: E-mail: mfeldt@mpia.de, Telephone: +49 6221 528 262}

\begin{document} 
\maketitle

\begin{abstract}
AOSAT is a python package for the analysis of single-conjugate adaptive optics (SCAO) simulation results. Python is widely used in the astronomical community these days, and AOSAT may be used stand-alone, integrated into a simulation environment, or can easily be extended according to a user's needs. Standalone operation requires the user to provide the residual wavefront frames provided by the SCAO simulation package used, the aperture mask (pupil) used for the simulation, and a custom setup file describing the simulation/analysis configuration. In its standard form, AOSAT's "tearsheet" functionality will then run all standard analyzers, providing an informative plot collection on properties such as the point-spread function (PSF) and its quality, residual tip-tilt, the impact of pupil fragmentation, residual optical aberration modes both static and dynamic, the expected high-contrast performance of suitable instrumentation with and without coronagraphs, and the power spectral density of residual wavefront errors. 

AOSAT fills the gap between the simple numerical outputs provided by most simulation packages, and the full-scale deployment of instrument simulators and data reduction suites operating on SCAO residual wavefronts. It enables instrument designers and end-users to quickly judge the impact of design or configuration decisions on the final performance of down-stream instrumentation.
\end{abstract}

\keywords{Adaptive Optics, Simulation, Design, Performance Evaluation}

\section{INTRODUCTION}
\label{sec:intro}  

Adaptive optics (AO), with its capabilities to correct optical disturbances caused by Earth's atmosphere in real time, is becoming ever more common in visible/infrared astronomy. Increasingly seen as a standard facility that supports any type of standard instrumentation, so-called single-conjugate AO (SCAO) is inherently part of a large number of next-generation projects currently in one of their pre-commissioning phases (METIS\cite{metis}; MICADO\cite{micado}; HARMONI\cite{harmoni}; NFIRAOS\cite{nfiraos}; GMT\cite{gmt}; for an overview see Hippler et al. 2019\cite{hippler:2019}). In these phases, the design of SCAO systems relies heavily on end-to-end simulations, the typical end-to-end (SC)AO simulation package such as YAO\cite{yao}, COMPASS\cite{compass}, OOMAO\cite{oomao}, or CAOS\cite{caos} provides a few numbers that characterize the system's performance during the simulation, plus the residual wavefronts at each time step. The numbers provided are typically the Strehl ratio, a single measure for the quality of an optical image, plus its variation across wavelengths, and according to separation from the AO reference star. For modern high-performance instrumentation these numbers are not sufficient to judge the performance effectively, and a deeper analysis of the residual wave fronts is required. AOSAT currently performs in-depth analyses of residual wavefronts focusing on different aspects such as (but not limited to) high-contrast imaging, the impact of fragmented pupils, and non-common path aberrations. AOSAT is on the one hand an integrated tool, capable of providing a summary "tearsheet" of the system performance from a given simulation output, on the other hand built in a modular fashion so that it can easily be extended with additional "analyzers" focusing on the user's area of interest. This paper describes the software, its purpose and design principle, how to obtain, install, use, and extend AOSAT, and the various analyses being performed in the current implementation.

\section{PURPOSE AND DESIGN}
\label{sec:purpose}

Many sophisticated (SC)AO simulation packages exist, and are used for a variety of purposes: Instrument development and design is a key use case, but fundamental research and application in laboratory setups is also among the applications of such software.  Most such packages produce a set of key numbers as output, the Strehl number being the most prominent one among these, plus the residual phases at each end-to-end simulation step.  Unsurprisingly, all of the information is in the residual phase data (getting them as close to zero as possible being the whole purpose of adaptive optics),  from which also all the numbers can be re-computed. 

Analyzing the residual phases is in most cases done by the same people who executed the simulation, often by means of custom, on-the-fly-written scripts.  This is sufficient in e.g. project work towards a specific goal, but it makes comparisons between projects difficult.  Different simulation software, different key metrics examined, and different definitions of {\textit key} metrics occur across the AO landscape, and custom tailored analysis solutions can routinely only comply to one specific set of requirements and definitions.

AOSAT is intended to provide a standardized platform that can analyze the output of many if not all SCAO simulation packages by feeding residual phases, the telescope pupil and a set-up file describing fundamental simulation parameters. On output, AOSAT provides a well-tested and standardized set of analyses with a well-documented set of metrics and graphical tools to have an insight in the AO correction quality.  AOSAT can be easily extended with custom-made additional analyzers. The "analyzer" object is an independent analysis tool focused on one particular aspect of the wavefront, of which many already exist in AOSAT, but more can be contributed.  Thanks to its modular architecture described below, AOSAT can be easily receive inputs (residual phase screens) from any AO simulator

\subsection{Design - Setup File \label{sec:setup}}

In addition to the residual phase screens and the telescope pupil, the main input required by AOSAT is the set-up file. This file must contain the parameters used for the AO simulation, as detailed in the documentation\cite{doc}.

To run AOSAT one needs to provide 
\begin{enumerate}
\item
phase screens generated by a proper end-to-end simulation tool
in \sphinxstyleabbreviation{FITS}\footnote{Flexible Image Transport System. Most simulation tools do or can easily produce this type of output which is simple-structured, human readable and
flawlessly implemented in just about every astronomical piece of software.
There is no plan to ever support HDF5.  If you want to know why, look at \url{https://cyrille.rossant.net/moving-away-hdf5/}}\cite{fits} files
in a subdirectory of your current working directory. 

\item
the telescope pupil.  It, too, must be stored in a FITS file, and the scale
in pixels per metre and its overall dimensions must naturally be the same as for the residual phase screens.
AOSAT operates with transmission numbers, so if your 2D pupil map contains a 1.0
at a given location, it is transparent, whereas if it contains a 0.0 it is opaque.
Intermediate values are welcome, i.e. the pupil map can also contain non-binary values representing transmission variations (e.g. the effect of segmented primary mirrors or apodizing masks).

\item
the set-up file describing key properties, such as the pixel scale, the unit in which residual phases are given, the simulation frequency, and the wavelength at which the analysis is desired.  In addition there are parameters that describe the organization of the input files, and certain switches that can be set according to the user's discretion.
\end{enumerate}

AOSAT can handle a multitude of file and frame ordering schemes, most of them simultaneously. Given a target directory and filename pattern

\begin{sphinxVerbatim}[commandchars=\\\{\}]
\PYG{n}{screen\PYGZus{}fpattern}  \PYG{o}{=} \PYG{l+s+s1}{\PYGZsq{}}\PYG{l+s+s1}{*residual\PYGZus{}phase*.fits}\PYG{l+s+s1}{\PYGZsq{}}
\end{sphinxVerbatim}

AOSAT will read all files matching the pattern.  They will be sorted in a somewhat intelligent way according to integer numbers found in the filenames. If certain or all individual files contain cubes of data along the NAXIS3 of the FITS file, such frames will be served sequentially one after the other as individual phase screens.
AOSAT relies on the phase screens having an equal temporal spacing throughout the sequence reflected in the \sphinxstyleliteralstrong{\sphinxupquote{loopfreq}} key in  the set-up file. Note that AOSAT is not inherently intended to analyze single phase screens, but most individual analyzers that do not focus on some temporal aspect can be set up to do so.

A typical full set-up file looks as follows:

\begin{sphinxVerbatim}[commandchars=\\\{\}]
\PYG{n}{pupilmask}        \PYG{o}{=} \PYG{l+s+s1}{\PYGZsq{}}\PYG{l+s+s1}{telescope\PYGZus{}pupil.fits}\PYG{l+s+s1}{\PYGZsq{}}
\PYG{n}{screen\PYGZus{}dir}       \PYG{o}{=} \PYG{l+s+s1}{\PYGZsq{}}\PYG{l+s+s1}{residual\PYGZus{}phasescreens}\PYG{l+s+s1}{\PYGZsq{}}
\PYG{n}{screen\PYGZus{}fpattern}  \PYG{o}{=} \PYG{l+s+s1}{\PYGZsq{}}\PYG{l+s+s1}{*residual\PYGZus{}phase*.fits}\PYG{l+s+s1}{\PYGZsq{}}
\PYG{n}{ppm}              \PYG{o}{=} \PYG{l+m+mf}{10.0} \PYG{c+c1}{\PYGZsh{} pixels pr metre}
\PYG{n}{loopfreq}         \PYG{o}{=} \PYG{l+m+mf}{1000.0} \PYG{c+c1}{\PYGZsh{} Hz \PYGZhy{} actually it\PYGZsq{}s the frequency of phase screens on disk}
\PYG{n}{phaseunit}        \PYG{o}{=} \PYG{l+s+s2}{\PYGZdq{}}\PYG{l+s+s2}{micron}\PYG{l+s+s2}{\PYGZdq{}}               \PYG{c+c1}{\PYGZsh{} unit of phase screens}
\PYG{n}{an\PYGZus{}lambda}        \PYG{o}{=} \PYG{l+m+mf}{3.0e\PYGZhy{}6}                 \PYG{c+c1}{\PYGZsh{} analysis wavelength in metres}
\end{sphinxVerbatim}

\subsection{Design - Serving Frames}

For flexibility, finding, sorting, reading, unit conversion and potentially zero-padding the input residual phases is decoupled from the actual analyses.  A so-called frameserver handles these tasks, and hands the residual phases one by one to the individual analyzers. In this way, not only can different output formats (single frame files, cubes of frames, flexible numbering scheme) be used with AOSAT, but it is also possible to replace the frameserver with a custom one that can e.g. read different file formats, or serve frames directly out of a simulation code. For quick test runs, the frameserver can also be directed to skip certain frames, or serve only a subset of a temporal sequence.

\subsection{Design - Analyzing Tools}

For reasons of flexibility, every analysis to be performed has its own analyzer object in AOSAT.  In general, these objects do not require dedicated parameters in the set-up file, other than if they are required to run or not. The choice to not require such parameters has been made in order to keep results comparable and reliable.  There are however a few exceptions to this rule (currently there is exactly one exception: The contrast analyzer can be set up to run either non-coronagraphic, or with an ideal coronagraph).

The deliberate design choice to keep all analyses independent of each other does cause some overheads, e.g. some FFTs are performed multiple times as there is no communication \textit{between} analyzers.  However it facilitates implementing new analyzers at ease. To do so, there is a base class \sphinxcode{\sphinxupquote{dmy\PYGZus{}analyzer}} from which new analyzers can be derived and its methods overwritten.  For details, see the documentation on readthedocs\cite{doc}.

During a run, each analyzer gets served the input residual phases one at a time, plus the information on how many frames to expect in total.  Once serving is finished, each analyzer provides a \sphinxcode{\sphinxupquote{finalize()}} method that needs to be called in order to perform concluding computations such as, of course, statistics over all frames.  After that, analyzers expose the results of the individual analysis in question via dedicated properties, and all internal ones provide the convenience methods \sphinxcode{\sphinxupquote{make\PYGZus{}report()}} and \sphinxcode{\sphinxupquote{make\PYGZus{}plot()}}.

For detailed description of available standard analyzers, and the way that analyses are actually performed, see Sec.~\ref{sec:analyzers}

\section{INSTALLING AOSAT}
\subsection{Installation}
\label{\detokenize{getting_started/installation:installing-while-still-in-development}}\begin{enumerate}
\sphinxsetlistlabels{\arabic}{enumi}{enumii}{}{.}%
\item {} 
Make a new empty directory, e.g. like this:

\end{enumerate}

\begin{sphinxVerbatim}[commandchars=\\\{\}]
\PYGZdl{} mkdir AOSAT
\PYGZdl{} cd AOSAT
\end{sphinxVerbatim}
\begin{enumerate}
\sphinxsetlistlabels{\arabic}{enumi}{enumii}{}{.}%
\setcounter{enumi}{1}
\item {} 
Clone the repository

\end{enumerate}

\begin{sphinxVerbatim}[commandchars=\\\{\}]
\PYGZdl{} git clone https://github.com/mfeldt/AOSAT.git
\end{sphinxVerbatim}
\phantomsection\label{\detokenize{getting_started/installation:virtualenv}}\begin{enumerate}
\sphinxsetlistlabels{\arabic}{enumi}{enumii}{}{.}%
\setcounter{enumi}{2}
\item {} 
Create and activate a virtual environment (if you don’t know/have virtualenv, use pip to install it)

\end{enumerate}

\begin{sphinxVerbatim}[commandchars=\\\{\}]
\PYGZdl{} virtualenv \PYGZhy{}p python3.6 venv
\PYGZdl{} source venv/bin/activate
\end{sphinxVerbatim}

(This assumes using bash, there’s also venv/bin/activate.csh and a few others)
\begin{enumerate}
\sphinxsetlistlabels{\arabic}{enumi}{enumii}{}{.}%
\setcounter{enumi}{3}
\item {} 
Change to the repository and install:

\end{enumerate}

\begin{sphinxVerbatim}[commandchars=\\\{\}]
\PYGZdl{} cd AOSAT
\PYGZdl{} python setup.py install
\end{sphinxVerbatim}

That’s it, python should install the package and all required dependencies!

\subsection{Verifying the Installation}
\label{\detokenize{getting_started/installation:verifying-the-installation}}
To verify that the installation is fine, you can do the following:
\begin{enumerate}
\sphinxsetlistlabels{\arabic}{enumi}{enumii}{}{.}%
\item {} 
run the test suite

\end{enumerate}

\begin{sphinxVerbatim}[commandchars=\\\{\}]
\PYGZdl{} python setup.py test
\end{sphinxVerbatim}
\begin{enumerate}
\sphinxsetlistlabels{\arabic}{enumi}{enumii}{}{.}%
\setcounter{enumi}{1}
\item {} 
Try the individual files

\end{enumerate}

\begin{sphinxVerbatim}[commandchars=\\\{\}]
\PYGZdl{} cd src/aosat
\PYGZdl{} python fftx.py
\PYGZdl{} python aosat\PYGZus{}cfg.py
\PYGZdl{} python util.py
\PYGZdl{} python analyze.py
\end{sphinxVerbatim}

Ideally everything should terminate without failures. Beware it may take a while.

\subsection{Documentation}

When following the installation instructions an examples directory is checked out along with the code. To find the examples, you can do the following in python (running of course, in
your activated AOSAT environment):

\begin{sphinxVerbatim}[commandchars=\\\{\}]
\PYG{g+gp}{\PYGZgt{}\PYGZgt{}\PYGZgt{} }\PYG{k+kn}{import} \PYG{n+nn}{os}
\PYG{g+gp}{\PYGZgt{}\PYGZgt{}\PYGZgt{} }\PYG{n}{os}\PYG{o}{.}\PYG{n}{path}\PYG{o}{.}\PYG{n}{join}\PYG{p}{(}\PYG{n}{os}\PYG{o}{.}\PYG{n}{path}\PYG{o}{.}\PYG{n}{split}\PYG{p}{(}\PYG{n}{aosat}\PYG{o}{.}\PYG{n+nv+vm}{\PYGZus{}\PYGZus{}file\PYGZus{}\PYGZus{}}\PYG{p}{)}\PYG{p}{[}\PYG{l+m+mi}{0}\PYG{p}{]}\PYG{p}{,}\PYG{l+s+s1}{\PYGZsq{}}\PYG{l+s+s1}{examples}\PYG{l+s+s1}{\PYGZsq{}}\PYG{p}{)}
\end{sphinxVerbatim}

In the examples directory, you will find a number of set-up files that can be adapted to your needs by changing the relevant parameters. You may also start with the minimum example given above in Sec.~\ref{sec:setup}. In order to do more than runnig the examples, the best way is of course to look at the documentation which is available from \url{https://aosat.readthedocs.io}. Here, you will find detailed descriptions of the setup files, the configuration of a particular analysis, and the outputs that AOSAT provides when a particular analysis is performed.

\section{STANDARD ANALYZERS}
\label{sec:analyzers}
\subsection{"Tearsheet" Functionality}

\label{sec:tearsheet}

   \begin{figure} [ht]
   \begin{center}
   \begin{tabular}{c} 
   \includegraphics[height=15cm]{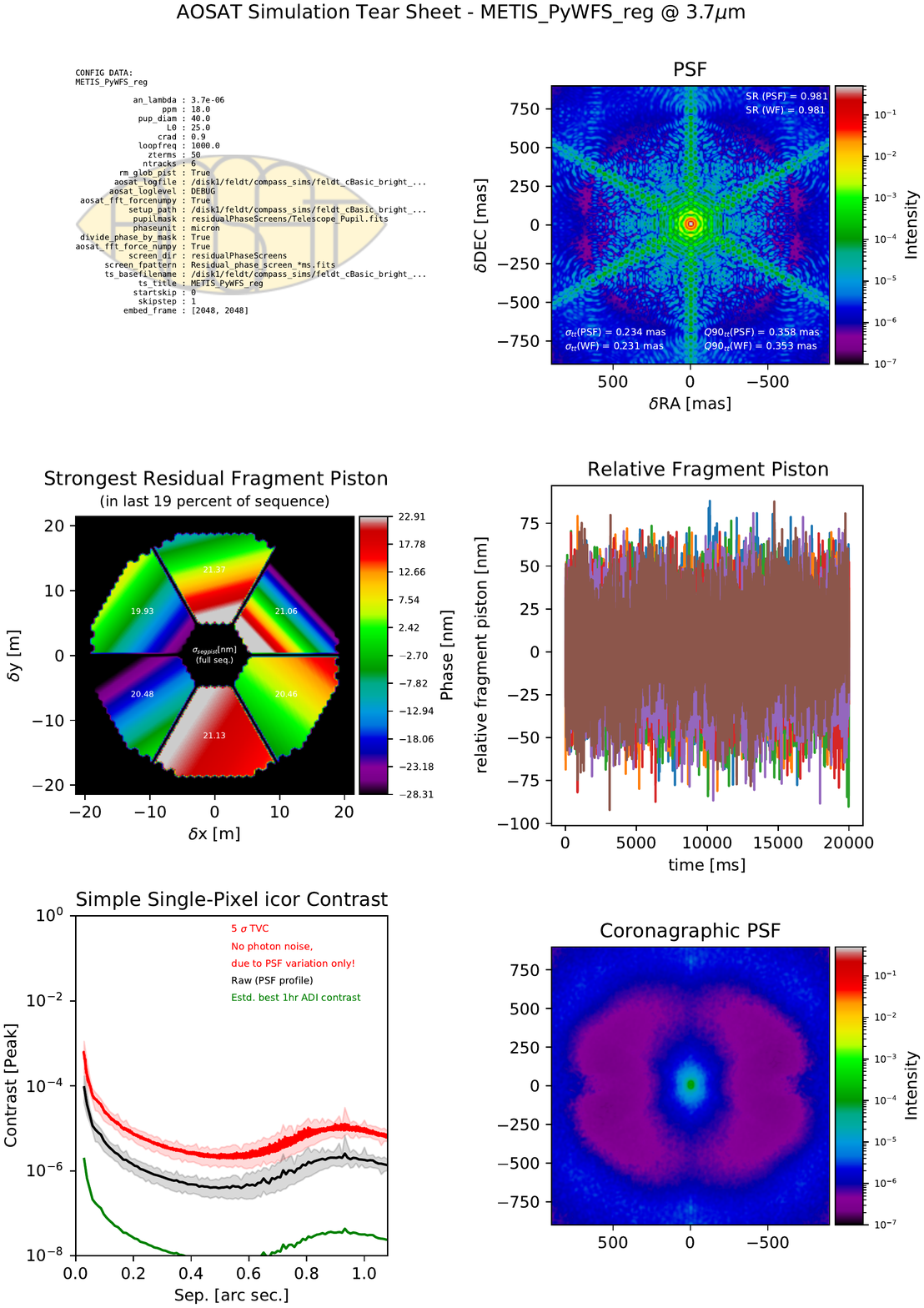}
   \end{tabular}
   \end{center}
   \caption[aosat_ts] 
   { \label{fig:aosat_ts} 
    Example tearsheet output of AOSAT.}
   \end{figure} 
   
A common use case is to produce a so\sphinxhyphen{}called “tearsheet”, a double\sphinxhyphen{}sided page
summarizing the most interesting performance indicators from the results of a given AO simulation.

This can be achieved easily for the provided close\sphinxhyphen{}loop example like so:

\begin{sphinxVerbatim}[commandchars=\\\{\}]
\PYG{g+gp}{\PYGZgt{}\PYGZgt{}\PYGZgt{} }\PYG{k+kn}{import} \PYG{n+nn}{aosat}
\PYG{g+gp}{\PYGZgt{}\PYGZgt{}\PYGZgt{} }\PYG{n}{example\PYGZus{}path} \PYG{o}{=} \PYG{n}{os}\PYG{o}{.}\PYG{n}{path}\PYG{o}{.}\PYG{n}{join}\PYG{p}{(}\PYG{n}{os}\PYG{o}{.}\PYG{n}{path}\PYG{o}{.}\PYG{n}{split}\PYG{p}{(}\PYG{n}{aosat}\PYG{o}{.}\PYG{n+nv+vm}{\PYGZus{}\PYGZus{}file\PYGZus{}\PYGZus{}}\PYG{p}{)}\PYG{p}{[}\PYG{l+m+mi}{0}\PYG{p}{]}\PYG{p}{,}\PYG{l+s+s1}{\PYGZsq{}}\PYG{l+s+s1}{examples}\PYG{l+s+s1}{\PYGZsq{}}\PYG{p}{)}
\PYG{g+gp}{\PYGZgt{}\PYGZgt{}\PYGZgt{} }\PYG{n}{example\PYGZus{}file} \PYG{o}{=} \PYG{n}{os}\PYG{o}{.}\PYG{n}{path}\PYG{o}{.}\PYG{n}{join}\PYG{p}{(}\PYG{n}{example\PYGZus{}path}\PYG{p}{,}\PYG{l+s+s1}{\PYGZsq{}}\PYG{l+s+s1}{example\PYGZus{}analyze\PYGZus{}closed\PYGZus{}loop.setup}\PYG{l+s+s1}{\PYGZsq{}}\PYG{p}{)}
\PYG{g+gp}{\PYGZgt{}\PYGZgt{}\PYGZgt{} }\PYG{n}{aosat}\PYG{o}{.}\PYG{n}{analyze}\PYG{o}{.}\PYG{n}{tearsheet}\PYG{p}{(}\PYG{n}{example\PYGZus{}file}\PYG{p}{)}
\end{sphinxVerbatim}

This will produce two files in the current working directory:

\begin{sphinxVerbatim}[commandchars=\\\{\}]
\PYG{n}{ts\PYGZus{}test}\PYG{o}{.}\PYG{n}{txt}
\PYG{n}{ts\PYGZus{}test}\PYG{o}{.}\PYG{n}{pdf}
\end{sphinxVerbatim}

Guess what the familiar extensions mean and look at them with the appropriate tool for each
to see what they are about.  An example can be seen in Fig.~\ref{fig:aosat_ts}.

\subsection{The PSF Analyzer}
\label{\detokenize{analyzers/psf_analyzer:description}}
Analyzing the \sphinxstyleabbreviation{PSF} (Point Spread Function) is done in a straight forward way: the residual phase of stored in each screen is transformed into a PSF by means of the well known:

\begin{sphinxVerbatim}[commandchars=\\\{\}]
\PYG{n}{np}\PYG{o}{.}\PYG{n}{abs}\PYG{p}{(}\PYG{n}{fftForward}\PYG{p}{(}\PYG{n}{aperture}\PYG{o}{*}\PYG{n}{np}\PYG{o}{.}\PYG{n}{exp}\PYG{p}{(}\PYG{l+m+mi}{1}\PYG{n}{j}\PYG{o}{*}\PYG{n}{phase}\PYG{p}{)}\PYG{p}{)}\PYG{p}{)}\PYG{o}{*}\PYG{o}{*}\PYG{l+m+mi}{2}
\end{sphinxVerbatim}

where \sphinxcode{\sphinxupquote{aperture}} represents the telescope entrance pupil defined in the setup file.
A number of additional performance indicators are calculated on both the resulting PSF, and the input phase.
This can e.g. reveal deviations from the Maréchal approximation when certain types of aberrations (e.g. waffle modes) occur. Such double computations are carried out for 

\begin{itemize}
\item {} 
the Strehl ratio is measured on the AO residual phase \(\phi\) as \(S = e^{-\sigma_\phi^2}\).

\item {} 
the Strehl ratio is measured on the PSF as \(S = I_{peak}/I_{ref, peak}\), where \(I\) is the PSF’s intensity distribution, and \(I_{ref}\) is the intensity distribution of a reference PSF resulting from a perfectly flat wavefront.

\item {} 
the tip and tilt excursion of the PSF is measured on the phase \(\phi\) by means of a least squares fit of a tilted flat wavefront to the phase

\item {} 
the tip and tilt excursion of the PSF is measured on the PSF by means of fitting a 2D Gaussian to the core of the PSF.

\end{itemize}

\subsubsection{Plot caption}

   \begin{figure} [ht]
   \begin{center}
   \begin{tabular}{c} 
   \includegraphics[width=0.5\textwidth]{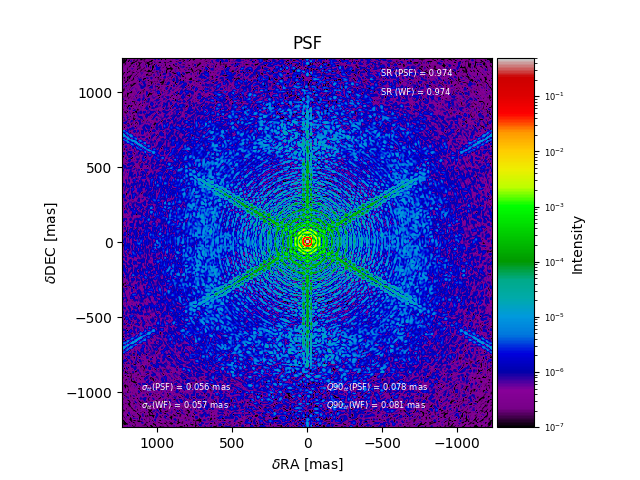}
   \end{tabular}
   \end{center}
   \caption[psf] 
   { \label{fig:psf} 
    Typical Figure produced by the PSF analyzer.}
   \end{figure}

\label{\detokenize{analyzers/psf_analyzer:plot-caption}}
When called on its own, or on a figure with sufficient available subplot space, \sphinxcode{\sphinxupquote{frg\_anaylzer.makeplot()}} will produce a figure as shown in Fig.~\ref{fig:psf}. The suggested caption for the figure could be:

\sphinxstyleemphasis{Resulting time\sphinxhyphen{}averaged PSF in units of peak intensity.  Additionally shown is the Strehl ratio derived from the peak intensity, denoted as “SR(PSF)”, and derived from the wavefront quality, denoted as “SR(WF)”. The tip\sphinxhyphen{}tilt statistics are shown in the lower part, also derived from the PSF directly as well as from the wavefronts.}  \(Q_{90}\) \sphinxstyleemphasis{means that 90\% of TT excursions are smaller than the quoted value.}

\subsubsection{Resulting properties}
\label{\detokenize{analyzers/psf_analyzer:resulting-properties}}
\sphinxtitleref{psf\_analyzer} exposes the following properties after \sphinxcode{\sphinxupquote{psf\_analyzer.finalize()}} has been called:

\begin{savenotes}\sphinxattablestart
\centering
\sphinxcapstartof{table}
\sphinxthecaptionisattop
\sphinxcaption{psf\_analyzer properties}\label{\detokenize{analyzers/psf_analyzer:id5}}
\sphinxaftertopcaption
\begin{tabular}[t]{|\X{2}{10}|\X{3}{10}|\X{5}{10}|}
\hline
\sphinxstyletheadfamily 
Property
&\sphinxstyletheadfamily 
type
&\sphinxstyletheadfamily 
Explanation
\\
\hline
\sphinxcrossref{\sphinxcode{\sphinxupquote{psf}}}
&
2D ndarray (float)
&
Time\sphinxhyphen{}averaged PSF.
\\
\hline
\sphinxcrossref{\sphinxcode{\sphinxupquote{strehl}}}
&
float
&
Strehl ratio of PSF derived from peak intensity.
\\
\hline
\sphinxcrossref{\sphinxcode{\sphinxupquote{sr\_wf}}}
&
float
&
Strehl ratio of PSF derived from residual wave fronts.
\\
\hline
\sphinxcrossref{\sphinxcode{\sphinxupquote{ttx}}}
&
1D ndarray (float) of length n\_frames
&
Global tip for each frame from Gauss\sphinxhyphen{}fitted PSF location (mas).
\\
\hline
\sphinxcrossref{\sphinxcode{\sphinxupquote{tty}}}
&
1D ndarray (float) of length n\_frames
&
Global tilt for each frame from Gauss\sphinxhyphen{}fitted PSF location (mas).
\\
\hline
\sphinxcrossref{\sphinxcode{\sphinxupquote{ttilt}}}
&
1D ndarray (float) of length n\_frames
&
Global excursion from centre, determined from wavefront (mas)
\\
\hline
\sphinxcrossref{\sphinxcode{\sphinxupquote{ttjit}}}
&
float
&
rms of \sphinxtitleref{ttilt}
\\
\hline
\sphinxcrossref{\sphinxcode{\sphinxupquote{ttq90}}}
&
float
&
90\% quantile of \sphinxtitleref{ttilt}
\\
\hline
\sphinxcrossref{\sphinxcode{\sphinxupquote{ttjit\_psf}}}
&
float
&
rms of $\sqrt{\textrm{ttx}^2+\textrm{tty}^2} $
\\
\hline
\sphinxcrossref{\sphinxcode{\sphinxupquote{ttq90\_psf}}}
&
float
&
90\% quantile of $\sqrt{\textrm{ttx}^2+\textrm{tty}^2} $
\\
\hline
\end{tabular}
\par
\sphinxattableend\end{savenotes}


\subsection{Pupil Fragmentation Analyzer}

\label{\detokenize{analyzers/frg_analyzer:description}}
The fragmentation analyzer looks, as the name implies, at pupil fragments individually.
Pupil fragments are sections of the telescope pupil that are not connected to one another and thus form individual
fragments or “islands”. Fragmentation may occur due to the secondary support structure (aka "spiders"), or by design of having several disconnected primary mirrors on a single support structure.  Spiders can cause the additional inconvenience of the low-wind effect (see below).

This fragmentation can give rise to two kinds of effects in adaptive optics, the
\begin{description}
\item[{Island effect}] \leavevmode
arises when the AO loop itself introduces independent piston terms for each fragment due to the reconstructor’s inability to provide accurate information on the wavefront’s piston offset across the spider. The other possibility is the

\item[{Low\sphinxhyphen{}wind effect}] \leavevmode
originally dubbed “Mickey Mouse effect” due to the shape of the PSF it produces, it occurs at very low wind speeds at ground level. Here, a physical phase difference between adjacent fragments exists due to the air being cooler on the downwind side of the radiatively cooled spider. This causes a phase jump across the spider, plus frequently a tilt across the affected downwind fragment as temperatures re-equilibrize further down the weak flow  (see e.g. Fig. 1 in \citenum{sauvage:2016}). Again combined with the reconstructor’s inability to yield accurate piston information, these do not get corrected.

\end{description}

The result of both is similar (apart from the possible fragment tilt caused by the low-wind effect), but it is important to note that the first is purely due to missing information, while the second has a physical origin.
Few AO simulations accurately simulate the low\sphinxhyphen{}wind effect correctly, it usually needs to be introduced specifically.

In any case tip\sphinxhyphen{}tilt terms remain closely connected to pupil fragmentation, as a global tilt across the pupil will also cause piston terms of fragments to differ, particularly along the tilt gradient.  Vice versa, a real piston difference of fragments opposing each other in the pupil (may) appear as a global tilt.

\subsubsection{Finding fragments}
\label{\detokenize{analyzers/frg_analyzer:finding-fragments}}
Analyzing pupil fragments is an integral part of AOSAT,  the fragments are thus identified during the setup of any analysis run, no matter which particular analyzer is used afterwards. Fragments are found by means of the \sphinxhref{https://docs.scipy.org/doc/scipy/reference/generated/scipy.ndimage.label.html}{scipy.ndimage.label} function, and are contained in the setup dictionary’s \sphinxcode{\sphinxupquote{fragmask}} element.

   \begin{figure} [ht]
   \begin{center}
   \begin{tabular}{c} 
   \includegraphics[width=0.5\textwidth]{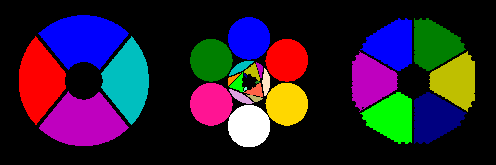}
   \end{tabular}
   \end{center}
   \caption[apertures] 
   { \label{fig:apertures} 
    Labeled apertures of some well\sphinxhyphen{}known telescopes / projects.}
   \end{figure}

\subsubsection{Analyzing fragments}
\label{\detokenize{analyzers/frg_analyzer:analyzing-fragments}}
At each time step, \sphinxcode{\sphinxupquote{frg\_analyzer}} determines the piston and tip\sphinxhyphen{}tilt terms of each individual fragment.   Piston and fragmental (i.e. global across the fragment) tip\sphinxhyphen{}tilt is determined by a least squares fit of a tilted but otherwise flat plane to the wavefront inside the fragment.

\begin{sphinxVerbatim}[commandchars=\\\{\}]
\PYG{k}{for} \PYG{n}{i} \PYG{o+ow}{in} \PYG{n+nb}{range}\PYG{p}{(}\PYG{n}{num\PYGZus{}fragments}\PYG{p}{)}\PYG{p}{:}
  \PYG{c+c1}{\PYGZsh{}\PYGZsh{} tilt from WF}
  \PYG{c+c1}{\PYGZsh{}\PYGZsh{} wfrag contains the valid indeces for each fragment}
  \PYG{n}{C}\PYG{p}{,}\PYG{n}{\PYGZus{}}\PYG{p}{,}\PYG{n}{\PYGZus{}}\PYG{p}{,}\PYG{n}{\PYGZus{}} \PYG{o}{=} \PYG{n}{np}\PYG{o}{.}\PYG{n}{linalg}\PYG{o}{.}\PYG{n}{lstsq}\PYG{p}{(}\PYG{n+nb+bp}{self}\PYG{o}{.}\PYG{n}{A}\PYG{p}{[}\PYG{n}{i}\PYG{p}{]}\PYG{p}{,} \PYG{n}{frame}\PYG{p}{[}\PYG{n+nb+bp}{self}\PYG{o}{.}\PYG{n}{wfrag}\PYG{p}{[}\PYG{n}{i}\PYG{p}{]}\PYG{p}{]}\PYG{p}{)}
  \PYG{n+nb+bp}{self}\PYG{o}{.}\PYG{n}{ttxt}\PYG{p}{[}\PYG{n+nb+bp}{self}\PYG{o}{.}\PYG{n}{ffed}\PYG{p}{,}\PYG{n}{i}\PYG{p}{]}    \PYG{o}{=} \PYG{n}{C}\PYG{p}{[}\PYG{l+m+mi}{0}\PYG{p}{]}
  \PYG{n+nb+bp}{self}\PYG{o}{.}\PYG{n}{ttyt}\PYG{p}{[}\PYG{n+nb+bp}{self}\PYG{o}{.}\PYG{n}{ffed}\PYG{p}{,}\PYG{n}{i}\PYG{p}{]}    \PYG{o}{=} \PYG{n}{C}\PYG{p}{[}\PYG{l+m+mi}{1}\PYG{p}{]}
  \PYG{n+nb+bp}{self}\PYG{o}{.}\PYG{n}{pistont}\PYG{p}{[}\PYG{n+nb+bp}{self}\PYG{o}{.}\PYG{n}{ffed}\PYG{p}{,}\PYG{n}{i}\PYG{p}{]} \PYG{o}{=} \PYG{n}{C}\PYG{p}{[}\PYG{l+m+mi}{2}\PYG{p}{]}
\end{sphinxVerbatim}

Piston, tip, and tilt of each fragment are stored for each time step.

Upon completion, i.e. when \sphinxtitleref{finalize()} is called, the analyzer computes the mean, and the standard deviation on each of the stored time series.

\subsubsection{Plot captions}

   \begin{figure} [ht]
   \begin{center}
   \begin{tabular}{c} 
   \includegraphics[width=1.0\textwidth]{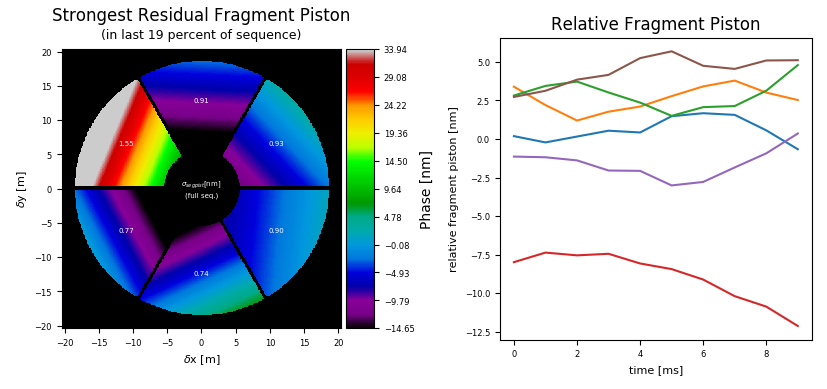}
   \end{tabular}
   \end{center}
   \caption[apertures] 
   { \label{fig:frg} 
    Plots produced by the pupil fragmentation analyzer.}
   \end{figure}

\label{\detokenize{analyzers/frg_analyzer:plot-captions}}
When called on its own, or on a figure with sufficient available subplot space, \sphinxcode{\sphinxupquote{frg\_anaylzer.makeplot()}} will produce two figures like as shown in Fig.~\ref{fig:frg}.
The figure caption for the left image would be:

\sphinxstyleemphasis{Pupil fragmentation  analysis.  The color image gives the piston and tilt of the frame with the largest span of piston values across fragments during the last
19\% (xx s) of the sequence. The numbers in the fragments give the piston value standard deviation (in nm) across the full sequence for the corresponding fragment.}

Note the length of the sequence to search for the worst piston occurrence can be altered by use of the \sphinxcode{\sphinxupquote{tile}} argument in the call of \sphinxcode{\sphinxupquote{frg\_analyzer.finalize(tile=0.8)}}.

The figure caption for the right image would be:

\sphinxstyleemphasis{Piston term of individual pupil fragments over time}

It is planned to provide the temporal power spectrum of piston terms as an inset in a forthcoming version.

\subsubsection{Resulting properties}
\label{\detokenize{analyzers/frg_analyzer:resulting-properties}}
\sphinxtitleref{frg\_analyzer} exposes the following properties after \sphinxcode{\sphinxupquote{frg\_analyzer.finalize()}} has been called:

\begin{savenotes}\sphinxattablestart
\centering
\sphinxcapstartof{table}
\sphinxthecaptionisattop
\sphinxcaption{frg\_analyzer porperties}\label{\detokenize{analyzers/frg_analyzer:id3}}
\sphinxaftertopcaption
\begin{tabular}[t]{|\X{2}{10}|\X{3}{10}|\X{5}{10}|}
\hline
\sphinxstyletheadfamily 
Property
&\sphinxstyletheadfamily 
type
&\sphinxstyletheadfamily 
Explanation
\\
\hline
\sphinxcrossref{\sphinxcode{\sphinxupquote{piston}}}
&
1D ndarray (float) of length n\_fragments
&
Array holding the mean piston value for each pupil fragment (in nm).
\\
\hline
\sphinxcrossref{\sphinxcode{\sphinxupquote{dpiston}}}
&
1D ndarray (float) of length n\_fragments
&
Array holding the standard deviation of the piston value for each pupil fragment (in nm).
\\
\hline
\sphinxcrossref{\sphinxcode{\sphinxupquote{pistont}}}
&
2D ndarray (float) of shape (n\_frames, n\_fragments).
&
Array holding individual piston values for each frame in the sequence.
\\
\hline
\sphinxcrossref{\sphinxcode{\sphinxupquote{ttx}}}
&
1D ndarray (float) of length n\_fragments
&
Array holding mean tip deviation for each fragment (in mas).
\\
\hline
\sphinxcrossref{\sphinxcode{\sphinxupquote{dttx}}}
&
1D ndarray (float) of length n\_fragments
&
Array holding the standard deviation of tip deviation for each pupil fragment (in mas).
\\
\hline
\sphinxcrossref{\sphinxcode{\sphinxupquote{ttxt}}}
&
2D ndarray (float) of shape (n\_frames, n\_fragments)
&
Array holding individual tip values for each frame in the sequence.
\\
\hline
\sphinxcrossref{\sphinxcode{\sphinxupquote{tty}}}
&
1D ndarray (float) of length n\_fragments
&
Array holding mean tilt deviation for each fragment (in mas).
\\
\hline
\sphinxcrossref{\sphinxcode{\sphinxupquote{dtty}}}
&
1D ndarray (float) of length n\_fragments
&
Array holding the standard deviation of tilt deviation for each pupil fragment (in mas).
\\
\hline
\sphinxcrossref{\sphinxcode{\sphinxupquote{ttyt}}}
&
2D ndarray (float) of shape (n\_frames, n\_fragments)
&
Array holding individual tilt values for each frame in the sequence.
\\
\hline
\sphinxcrossref{\sphinxcode{\sphinxupquote{pistframe}}}
&
2D ndarray (float)
&
Frame holding the worst piston pattern across the top (1\sphinxhyphen{}attr:\sphinxtitleref{tile}) (see below) part of the simulated sequence.
\\
\hline
\sphinxcrossref{\sphinxcode{\sphinxupquote{tile}}}
&
float
&
Fractional tile above which the analyzer looks for the worst piston frame in the sequence.
\\
\hline
\end{tabular}
\par
\sphinxattableend\end{savenotes}

\subsection{Zernike Expansion}

\label{\detokenize{analyzers/zrn_analyzer:description}}
\sphinxcode{\sphinxupquote{zrn\_analyzer}} yields the time averaged Zernike expansion of the individual residual phase frames.
The average, and the standard deviation of each term are calculated and presented.

\paragraph{Basis generation}
\label{\detokenize{analyzers/zrn_analyzer:basis-generation}}
The number of Zernike terms each residual wavefront is expanded into is determined by the \sphinxcode{\sphinxupquote{zterms}} key in the setup file.

The function called to set up the basis is \sphinxtitleref{poppy.zernike.arbitrary\_basis() from the poppy package\cite{poppy}}.  If you want to set up your own basis to expand wavefronts into, you should create a numpy (cupy) array of shape (d,d,nterms), where \(d\) is the pupil diameter of your aperture array in \sphinxcode{\sphinxupquote{sd{[}\textquotesingle{}tel\_mirror\textquotesingle{}{]}}} (see \sphinxhref{../general\_concept/setup}{setup file}).  Each plane needs to contain a phase map of the corresponding basis term.  All non\sphinxhyphen{}zero pixels in the aperture must be covered by the basis, else the analyzer will crash.  This array should be inserted into the setup dictionary key \sphinxcode{\sphinxupquote{zernike\_basis}}.

\begin{sphinxVerbatim}[commandchars=\\\{\}]
\PYG{n}{sd} \PYG{o}{=} \PYG{n}{aosat}\PYG{o}{.}\PYG{n}{analyze}\PYG{o}{.}\PYG{n}{setup}\PYG{p}{(}\PYG{p}{)}
\PYG{n}{sd}\PYG{p}{[}\PYG{l+s+s1}{\PYGZsq{}}\PYG{l+s+s1}{zernike\PYGZus{}basis}\PYG{l+s+s1}{\PYGZsq{}}\PYG{p}{]} \PYG{o}{=} \PYG{n}{my\PYGZus{}basis\PYGZus{}array}
\PYG{n}{a} \PYG{o}{=} \PYG{n}{aosat}\PYG{o}{.}\PYG{n}{analyze}\PYG{o}{.}\PYG{n}{zrn\PYGZus{}analyzer}\PYG{p}{(}\PYG{n}{sd}\PYG{p}{)}
\end{sphinxVerbatim}

Note the order of these statements which is crucial for the result to be as expected.
Of course your own basis does not necessarily need to be a basis of type Zernike.

\paragraph{Wavefront expansion}
\label{\detokenize{analyzers/zrn_analyzer:wavefront-expansion}}
The expansion of individual wavefronts itself is done by \sphinxcrossref{\sphinxcode{\sphinxupquote{aosat.util.basis\_expand()}}}, which computes the cross\sphinxhyphen{}correlations between the wavefront and individual basis terms.

\subsubsection{Plot captions}

   \begin{figure} [ht]
   \begin{center}
   \begin{tabular}{c} 
   \includegraphics[width=0.5\textwidth]{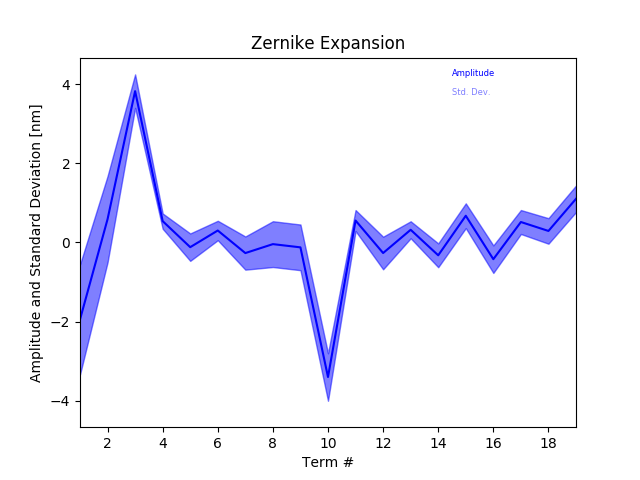}
   \end{tabular}
   \end{center}
   \caption[apertures] 
   { \label{fig:zrn} 
    Plot produced by the Zernike analyzer.}
   \end{figure} 

\label{\detokenize{analyzers/zrn_analyzer:plot-captions}}
When called on its own, or on a figure with sufficient available subplot space, \sphinxcode{\sphinxupquote{aosat.analyze.frg\_anaylzer.makeplot()}} will produce a figure like shown in Fig.~\ref{fig:zrn}. The caption would be:

\sphinxstyleemphasis{Time\sphinxhyphen{}averaged Zernike expansion of residual wavefronts. The blue line denotes the average amplitude of each term,
the shaded area ranges from the average plus one standard deviation to the average minus one standard deviation.}

Reading this section you have probably noticed that the zrn\_analzer is somewhat mis\sphinxhyphen{}labeled as “Zernike” analyzer, since the functionalities to create the basis, and to expand the wavefront is actually outside of the analyzer. The only core functionality currently is to restrict the basis choice to Zernike, and produce the plot/report.

\subsubsection{Resulting Properties}
\label{\detokenize{analyzers/zrn_analyzer:resulting-properties}}
\sphinxcrossref{\sphinxcode{\sphinxupquote{aosat.analyzers\_.zrn\_analyzer}}} exposes the following properties after \sphinxcode{\sphinxupquote{aosat.analyzers\_.zrn\_analyzer.finalize()}} has been called:

\begin{savenotes}\sphinxattablestart
\centering
\sphinxcapstartof{table}
\sphinxthecaptionisattop
\sphinxcaption{zrn\_analyzer properties}\label{\detokenize{analyzers/zrn_analyzer:id2}}
\sphinxaftertopcaption
\begin{tabular}[t]{|\X{1}{9}|\X{3}{9}|\X{5}{9}|}
\hline
\sphinxstyletheadfamily 
Property
&\sphinxstyletheadfamily 
type
&\sphinxstyletheadfamily 
Explanation
\\
\hline
\sphinxcrossref{\sphinxcode{\sphinxupquote{modes}}}
&
1D float NDarray of length zterms
&
Time averaged mean amplitude of each mode
\\
\hline
\sphinxcrossref{\sphinxcode{\sphinxupquote{dmodes}}}
&
1D float NDarray of length zterms
&
Standard deviation of each mode amplitude
\\
\hline
\sphinxcrossref{\sphinxcode{\sphinxupquote{modest}}}
&
2D float NDarray of shape (zterms, nframes)
&
Individual modal expansion for each frame
\\
\hline
\end{tabular}
\par
\sphinxattableend\end{savenotes}

\subsection{Residual Phase}

\sphinxcrossref{\sphinxcode{\sphinxupquote{phs\_analyzer}}} is actually more a decorative displayer of the last residual phase frame than an actual analyzer.  Nevertheless, it tracks the rms of each input phase frame and computes the mean rms in the end.

\subsubsection{Plot caption}

   \begin{figure} [ht]
   \begin{center}
   \begin{tabular}{c} 
   \includegraphics[width=0.5\textwidth]{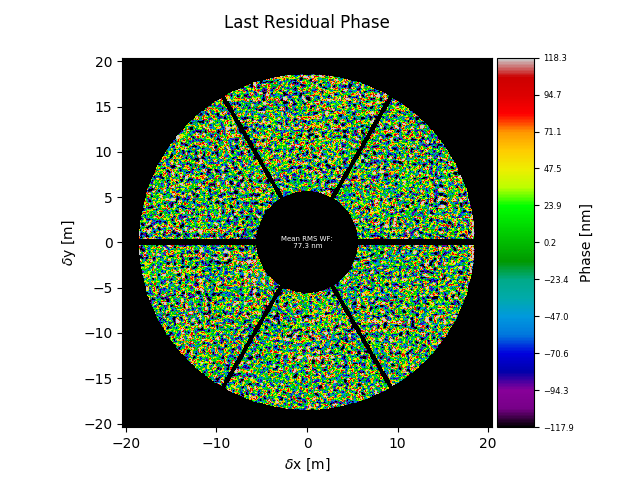}
   \end{tabular}
   \end{center}
   \caption[apertures] 
   { \label{fig:phs} 
    Plot produced by the residual phase analyzer.}
   \end{figure} 

\label{\detokenize{analyzers/phs_analyzer:plot-caption}}
When called on its own after \sphinxcode{\sphinxupquote{aosat.analyze.phs\_analyze.finalize()}}, \sphinxcode{\sphinxupquote{aosat.analyze.phs\_analyze.make\_plot()}} will produce a plot as shown in Fig.~\ref{fig:phs}.
The caption would read:

\sphinxstyleemphasis{Last residual wavefront in nm.  The number in the center gives the time\sphinxhyphen{}averaged rms of the residual wavefront.}

\subsubsection{Resulting properties}
\label{\detokenize{analyzers/phs_analyzer:resulting-properties}}
After calling \sphinxcode{\sphinxupquote{aosat.analyze.phs\_analyze.finalize()}}, phs\_analyzer will expose the following properties:

\begin{savenotes}\sphinxattablestart
\centering
\sphinxcapstartof{table}
\sphinxthecaptionisattop
\sphinxcaption{phs\_analyzer properties}\label{\detokenize{analyzers/phs_analyzer:id1}}
\sphinxaftertopcaption
\begin{tabular}[t]{|\X{2}{10}|\X{3}{10}|\X{5}{10}|}
\hline
\sphinxstyletheadfamily 
Property
&\sphinxstyletheadfamily 
type
&\sphinxstyletheadfamily 
Explanation
\\
\hline
\sphinxcrossref{\sphinxcode{\sphinxupquote{rms}}}
&
float
&
mean RMS of all wavefronts in nm.
\\
\hline
\sphinxcrossref{\sphinxcode{\sphinxupquote{rmst}}}
&
1D NDarray of length n\_frames
&
individual rms of all residual phase frames in nm
\\
\hline
\sphinxcrossref{\sphinxcode{\sphinxupquote{lastphase}}}
&
2D array
&
Last residual phasescreen (in nm)
\\
\hline
\end{tabular}
\par
\sphinxattableend\end{savenotes}

\subsection{Temporal Variance Contrast}

\label{\detokenize{analyzers/tvc_analyzer:description}}

\subsubsection{Motivation}
\label{\detokenize{analyzers/tvc_analyzer:motivation}}
High\sphinxhyphen{}contrast performance is one of the prime metrics to judge the quality of closed\sphinxhyphen{}loop operation of AO systems these days.  Thus, a contrast curve must not be missing from any decent analysis tearsheet for a given AO simulation. In addition, a kind of quick-look at focal plane residuals after removing the part of the PSF which is purely due to entrance pupil diffraction can reveal ultimate contrast limitations and underlying causes such as the actuator geometry or a wind-driven halo\cite{cantalloube:2020}.

The prime motivation to come up with \sphinxstyleabbreviation{TVC} (Temporal Variance Contrast) was the following list of requirements:
Computing the TVC should allow to
\begin{enumerate}
\sphinxsetlistlabels{\arabic}{enumi}{enumii}{}{.}%
\item {} 
be able to quickly access the contrast of a large set of simulations (hundreds) that produced a large number of residual WF frames (thousands) each.

\item {} 
not have to rely on specific assumptions about position in the sky, amount of field rotation, \sphinxstyleabbreviation{ADI} (Angular Differential Imaging) algorithm used, etc.

\item {} 
assess essentially the impact of a given deterioration of wavefront quality, rather than predict the precise contrast value.

\end{enumerate}

No. 1 inhibits the use of a full-fledged analysis software like such as VIP\cite{GomezGonzalez:2017} (\url{https://github.com/vortex-exoplanet/VIP}) coupled to a full model of the wavefront propagation through the coronagraphic optical train as realized by the HEEPS\cite{heeps} package., which usually runs a couple of hours on each simulation output. We needed something much simpler and faster.

No. 2 fostered the idea of coming up with a measure that attempts a measurement of those limitations that any angular differential imaging (ADI)\cite{marois:2006} algorithm would be unable to overcome. This has only partly been achieved, but the underlying thoughts are: Static aberrations cause static PSF patterns, which are easy to handle and cause little or no noise. Aberrations that are de\sphinxhyphen{}correlated between frames cause independent realizations of random PSF patterns, these can essentially not be overcome by ADI methods (But instead average out to a certain degree over time). That leaves aberrations that change in intermediate time\sphinxhyphen{}scales for the ADI algorithms to handle. These, however, are hardly present in many simulations as SCAO simulation packages ususally do not simulate \sphinxstyleabbreviation{NCPA} (Non\sphinxhyphen{}Common Path Aberrations) variations, flexure, pupil shifts, or any other slowly-changing  effect yet.

So in order to measure the fundamental limit imposed by the independent realizations, we came up with the idea of looking at the variance along the temporal axis at each image location separately. This is nicely independent of any assumed amount of field rotation, which would not change these statistics. As variance can be computed in an on\sphinxhyphen{}line fashion on focal\sphinxhyphen{}plane frames being computed one\sphinxhyphen{}by\sphinxhyphen{}one subsequently, the method is suitable to operate on very large data sets that cannot be held in memory entirely. Since we are not heavily interested in the absolute contrast value, we operate on single pixels only rather than averaging over a certain aperture.

\subsubsection{Definition}
\label{sec:tvc:def}
For this section, we compute a comparison between TVC and a standard ADI contrast. We use a simulation data set from the \sphinxstyleabbreviation{METIS} (Mid\sphinxhyphen{}Infrared ELT Imager and Spectrograph)\cite{metis} preliminary design phase.

We define two types of contrast to compare, in addition to the actual HCI\footnote{High-Contrast Imaging, using a full-fledged ADI package to analyze the simulated data.} result. First the TVC. It is computed in the following way:
\begin{enumerate}
\sphinxsetlistlabels{\arabic}{enumi}{enumii}{}{.}%
\item {} 
Compute image frames from the residual phases and stack in a cube

\item {} 
Compute the variance \(V\) along the temporal axis for each pixel

\item {} 
For a given separation, average the variances over an annulus covering that separation

\item {} 
Compute the 5$\sigma$ contrast for that separation as \(5 \times \sqrt{\bar{V}}\)

\end{enumerate}

In order to have a simplified\footnote{The main simplification here is of course the absence of any actual field rotation} model for ADI, we compute what we will call the ADI contrast in the following way
\begin{enumerate}
\sphinxsetlistlabels{\arabic}{enumi}{enumii}{}{.}%
\item {} 
Compute image frames from the residual phases and stack in a cube

\item {} 
Compute a robust mean image along the temporal axis, robust meaning we exclude the 10\% of values furthest from the median in each temporal vector.

\item {} 
Subtract the above image from each frame

\item {} 
Average along the temporal axis

\item {} 
For a given separation, compute the variance \(V\) of pixel values over an annulus covering that separation

\item {} 
Compute the 5$\sigma$ contrast for that separation as \(5 \times \sqrt{V}\)

\end{enumerate}

Note that when the assumption works that the temporal variance is a good measure for the variations that cannot be overcome by the ADI algorithm, the TVC contrast and the ADI contrast computed in this way should be related by the square root of the number frames. This is because the ADI procedure averages along the temporal axis, and the error of the mean (found in the spatial standard deviation when ADI contrast is measured) should be given by the temporal standard
deviation divided by the square root of the number of independent realizations.

\subsubsection{Comparison to ADI}
\label{\detokenize{analyzers/tvc_analyzer:comparison-to-adi}}

In order to compare the various methods to measure contrast, we computed the TVC and the above simplified ADI model on the same residual phase cube that was used to derive the contrasts in the METIS \sphinxstyleabbreviation{PDR} (Preliminary Design Review).

   \begin{figure} [ht]
   \begin{center}
   \begin{tabular}{c} 
   \includegraphics[width=0.5\textwidth]{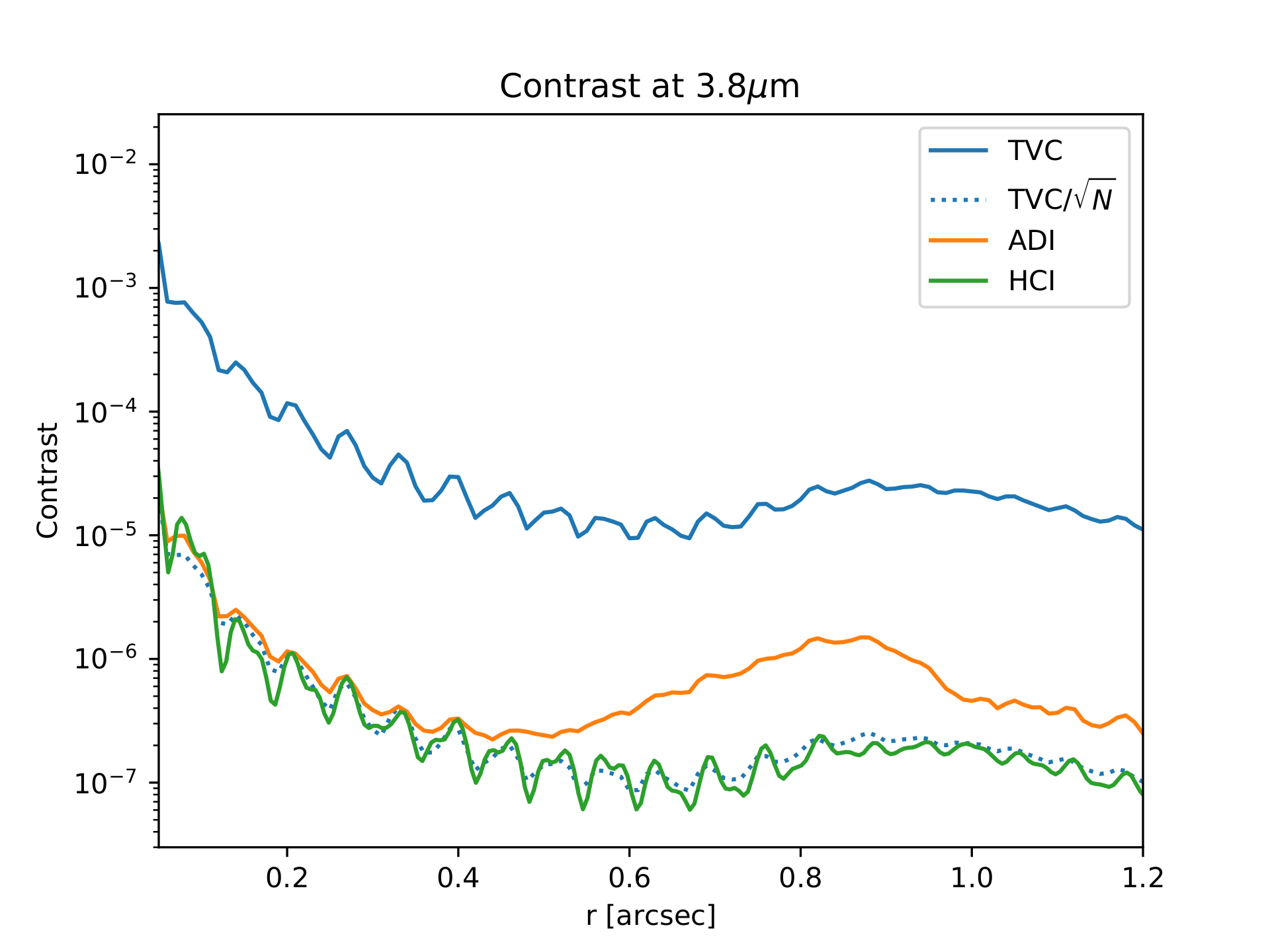}
   \end{tabular}
   \end{center}
   \caption[curves] 
   { \label{fig:ccc} 
    Contrast curves measured on METIS'\cite{metis} standard HCI data set. The green curve is the actual ADI result obtained by running VIP\cite{GomezGonzalez:2017} on the resulting image cube.
The straight blue line represents the ADI contrast measured with the method defined in Sec.~\ref{sec:tvc:def}. The dotted blue line is the same divided by the square root of the number of frames in the cube. The orange curve represents our simplified ADI model, also defined in Sec.\ref{sec:tvc:def}}
   \end{figure}

Fig~\ref{fig:ccc} shows the contrast curves derived in the various ways. Two observations can be made. Firstly, the match of the TVC curve divided by the square root of the number of frames to the contrast curve measured by true ADI via the VIP package is excellent. This said, we have to add two notes of caution here:

Firstly, we are working on simulated data and the detailed comparison has been made on one particular instance only. More in-depth analysis is needed, and ultimately it may be of great interest to perform a similar comparison on real data from a high-contrast imaging instrument.

Secondly, the input for this experiment is sampled at 300 ms steps, all frames are completely de\sphinxhyphen{}correlated from one another. For continuous simulation data sampled at much higher frequency, 
\sphinxcode{\sphinxupquote{tvc\_analyzer}} tries to determine the correlation length of a given input cube, and divide by the number of independent realizations instead of the number of frames in order to predict final 1hr ADI contrast. If this determination goes wrong for any reason, results will be unreliable!

The temporal statistics do not vary with time in these simulations \sphinxhyphen{} the TVC contrast measured on 500 frames is the same as the one measured on the full set
of 12,000 frames. Thus, if the condition of independent frames and no mid\sphinxhyphen{}temporal\sphinxhyphen{}frequency being present in the system is granted, the analysis can be greatly accelerated by running only on a subset!

Secondly, the simplified ADI model matches the scaled TVC and the measured ADI curve only partly. While the match in the interesting region around \(5\lambda/D\) is reasonable, the curves diverge around the control radius. It is beyond the scope of this comparison to investigate the details of this behaviour.

\subsubsection{Impact of wavefront quality}
\label{sec:tvc:imp}

   \begin{figure} [ht]
   \begin{center}
   \begin{tabular}{c} 
   \includegraphics[width=0.5\textwidth]{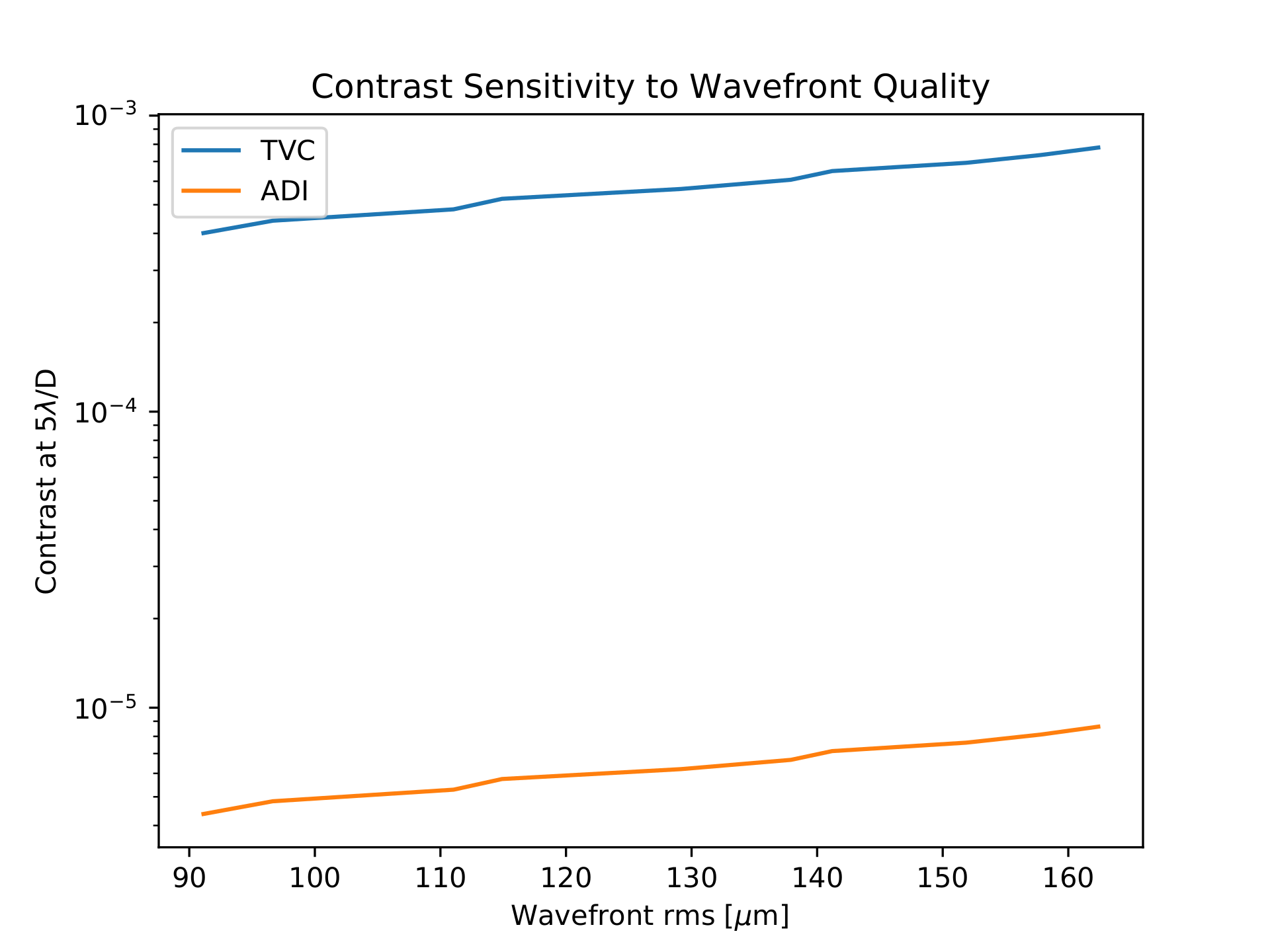}
   \end{tabular}
   \end{center}
   \caption[impact] 
   { \label{fig:csc} 
    Evolution of contrast versus wavefront quality.}
   \end{figure}

While the prediction of the actual high\sphinxhyphen{}contrast performance is nice\sphinxhyphen{}to\sphinxhyphen{}have, the prime goal of a contrast analysis in AO simulations is to catch all factors that have a pronounced impact on the high\sphinxhyphen{}contrast performance of the system.

In order to investigate this, we deteriorated the wavefronts by multiplying residual phase screens with a factor. The impact on contrast at 5$\lambda$/D can be seen in Fig.~\ref{fig:csc}. 
TVC and modeled ADI contrast as defined in Sec.~\ref{sec:tvc:def} behave nicely in parallel. 
TVC appears to be slightly less impacted than the ADI model, but not to a level larger than the usual uncertainties of contrast loss predictions.  We conclude, that we can safely use our TVC analyses to find critical impacts on contrast with this method.

\subsubsection{tvc\_analyzer}
\label{\detokenize{analyzers/tvc_analyzer:tvc-analyzer}}
\sphinxcode{\sphinxupquote{tvc\_analyzer}} is different from most other analyzers in AOSAT as it can run in two distinct mode: Coronagraphic and non\sphinxhyphen{}coronagraphic.  This is selected during instantiation by means of the \sphinxcode{\sphinxupquote{ctype}} keyword:

\begin{sphinxVerbatim}[commandchars=\\\{\}]
\PYG{n}{a} \PYG{o}{=} \PYG{n}{tvc\PYGZus{}analyzer}\PYG{p}{(}\PYG{n}{sd,ctype}\PYG{o}{=}\PYG{l+s+s1}{\PYGZsq{}}\PYG{l+s+s1}{icor}\PYG{l+s+s1}{\PYGZsq{}}\PYG{p}{)} \PYG{c+c1}{\PYGZsh{} run with an ideal coronagraph inserted}
\PYG{n}{a} \PYG{o}{=} \PYG{n}{tvc\PYGZus{}analyzer}\PYG{p}{(}\PYG{n}{sd, ctype}\PYG{o}{=}\PYG{l+s+s1}{\PYGZsq{}}\PYG{l+s+s1}{nocor}\PYG{l+s+s1}{\PYGZsq{}}\PYG{p}{)} \PYG{c+c1}{\PYGZsh{} run without coronagraph}
\PYG{n}{a} \PYG{o}{=} \PYG{n}{tvc\PYGZus{}analyzer}\PYG{p}{(}\PYG{n}{sd}\PYG{p}{)}              \PYG{c+c1}{\PYGZsh{} run without coronagraph}
\end{sphinxVerbatim}

When running in coronagraphic mode, the PSF creation from each residual phase frame is routed through a ideal coronagraph as described in Cavarroc et al. (2005)\cite{cavarroc:2005}.
In this case, the incoming complex amplitude \(A\) (represented as \sphinxcode{\sphinxupquote{tel\_mirror * exp(1j*phase)}}) is modified to \(\bar{A} = A - \Pi\), where \(\Pi\) represents the telescope pupil.  The usual factor of a square root of Strehl \(\sqrt{S}\) is not implemented in \sphinxcode{\sphinxupquote{tvc\_analyzer}}, as Strehl is either so high that it’s negligible, or the determination of \(S\) is unreliable.  Thus the implementation is \sphinxcode{\sphinxupquote{tel\_mirror * exp(1j*phase) \sphinxhyphen{} tel\_mirror}}.

In addition, \sphinxcode{\sphinxupquote{tvc\_analyzer}} produces an additional plot by default when instantiated in coronagraphic mode:  The ideal coronagraphic PSF.

\subsubsection{Plot captions}

   \begin{figure} [ht]
   \begin{center}
   \begin{tabular}{c} 
   \includegraphics[width=0.5\textwidth]{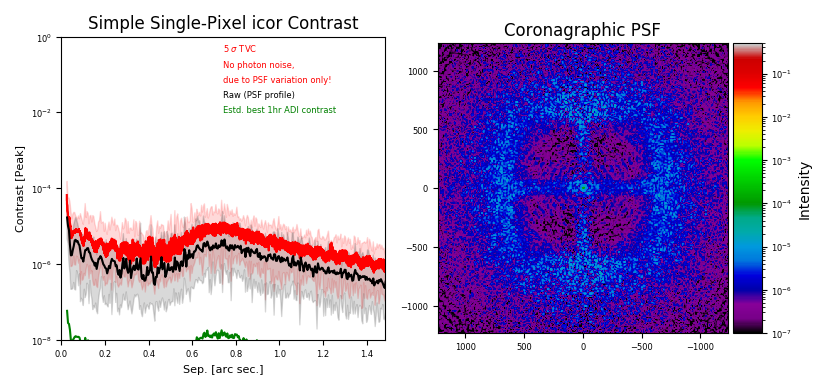}
   \end{tabular}
   \end{center}
   \caption[tvc] 
   { \label{fig:tvc} 
    Plots produced by the temporal variance contrast analyzer.}
   \end{figure} 

\label{\detokenize{analyzers/tvc_analyzer:plot-captions}}
When called on its own in coronagraphic mode, or on a figure with sufficient available subplot space, \sphinxcode{\sphinxupquote{tvc\_anaylzer.makeplot()}} will produce two figures like shown in Fig.~\ref{fig:tvc}

The figure caption for the left image would be :

\sphinxstyleemphasis{Resulting contrast curves.  The black curve shows the PSF profile, the red curve the resulting $5\sigma$ contrast from temporal variation of the PSF.
The green curve shows the predicted contrast limit achievable in an integration of 1\,hr after ADI processing.}

The figure caption for the right image would be :

\sphinxstyleemphasis{Time\sphinxhyphen{}averaged coronagraphic PSF. Intensities are relative to the peak of the non\sphinxhyphen{}coronagraphic PSF.}

In the non\sphinxhyphen{}coronagraphic case,  the right figure is missing.  The green curve is plotted only if a successful determination of the correlation time could be achieved.

\subsubsection{Resulting properties}
\label{\detokenize{analyzers/tvc_analyzer:resulting-properties}}
\sphinxtitleref{tvc\_analyzer} exposes the following properties after \sphinxcrossref{\sphinxcode{\sphinxupquote{finalize()}}} has been called:

\begin{savenotes}\sphinxattablestart
\centering
\sphinxcapstartof{table}
\sphinxthecaptionisattop
\sphinxcaption{tvc\_analyzer properties}\label{\detokenize{analyzers/tvc_analyzer:id7}}
\sphinxaftertopcaption
\begin{tabular}[t]{|\X{2}{9}|\X{2}{9}|\X{5}{9}|}
\hline
\sphinxstyletheadfamily 
Property
&\sphinxstyletheadfamily 
type
&\sphinxstyletheadfamily 
Explanation
\\
\hline
\sphinxcrossref{\sphinxcode{\sphinxupquote{ctype}}}
&
str
&
type of coronagraph (“icor” or “nocor”)
\\
\hline
\sphinxcrossref{\sphinxcode{\sphinxupquote{mean}}}
&
2D ndarray
&
time\sphinxhyphen{}averaged PSF.
\\
\hline
\sphinxcrossref{\sphinxcode{\sphinxupquote{variance2}}}
&
2D ndarray
&
variance2{[}1{]} contains the non\sphinxhyphen{}coronagraphic time\sphinxhyphen{}averaged PSF (\sphinxtitleref{icor} only)
\\
\hline
\sphinxcrossref{\sphinxcode{\sphinxupquote{contrast}}}
&
2D ndarray
&
5 sigma contrast of the PSF
\\
\hline
\sphinxcrossref{\sphinxcode{\sphinxupquote{rcontrast}}}
&
2D ndarray
&
raw contrast, i.e. the normalized PSF.
\\
\hline
\sphinxcrossref{\sphinxcode{\sphinxupquote{rvec}}}
&
2D ndarray (float)
&
Image where each pixel contains distance to centre (mas)
\\
\hline
\sphinxcrossref{\sphinxcode{\sphinxupquote{cvecmean}}}
&
2D ndarray (float)
&
Mean TVC contrast at locations in rvec.
\\
\hline
\sphinxcrossref{\sphinxcode{\sphinxupquote{cvecmin}}}
&
2D ndarray (float)
&
Minimum TVC contrast at locations in rvec.
\\
\hline
\sphinxcrossref{\sphinxcode{\sphinxupquote{cvecmax}}}
&
2D ndarray (float)
&
Maximum TVC contrast at locations in rvec.
\\
\hline
\sphinxcrossref{\sphinxcode{\sphinxupquote{rvecmean}}}
&
2D ndarray (float)
&
Mean raw contrast at locations in rvec.
\\
\hline
\sphinxcrossref{\sphinxcode{\sphinxupquote{rvecmin}}}
&
2D ndarray (float)
&
Minimum raw contrast at locations in rvec.
\\
\hline
\sphinxcrossref{\sphinxcode{\sphinxupquote{rvecmax}}}
&
2D ndarray (float)
&
Maximum raw contrast at locations in rvec.
\\
\hline
\sphinxcrossref{\sphinxcode{\sphinxupquote{corrlen}}}
&
float
&
Measured correlation length {[}\#frames{]}
\\
\hline
\sphinxcrossref{\sphinxcode{\sphinxupquote{max\_no\_cor}}}
&
float
&
Peak intensity of non\sphinxhyphen{}coronagraphic PSF
\\
\hline
\end{tabular}
\par
\sphinxattableend\end{savenotes}


\subsection{Spatial Power Spectrum}
The power spectrum analyzer derives the time\sphinxhyphen{}averaged spatial power spectrum of the residual phase frames.

\subsubsection{Anti\sphinxhyphen{}aliasing}
\label{\detokenize{analyzers/sps_analyzer:anti-aliasing}}
The spatial power spectrum is derived by an FFT of the input residual phase:

\begin{sphinxVerbatim}[commandchars=\\\{\}]
\PYG{n}{fftArray}       \PYG{o}{=} \PYG{n}{np}\PYG{o}{.}\PYG{n}{fft}\PYG{o}{.}\PYG{n}{fftshift}\PYG{p}{(}\PYG{n}{np}\PYG{o}{.}\PYG{n}{fft}\PYG{o}{.}\PYG{n}{fft2}\PYG{p}{(}\PYG{n}{ps}\PYG{o}{*}\PYG{n}{mask}\PYG{p}{,}\PYG{n}{norm}\PYG{o}{=}\PYG{l+s+s1}{\PYGZsq{}}\PYG{l+s+s1}{ortho}\PYG{l+s+s1}{\PYGZsq{}}\PYG{p}{)}\PYG{p}{)}
\PYG{n}{fftArray}       \PYG{o}{=} \PYG{p}{(}\PYG{n}{fftArray} \PYG{o}{*} \PYG{n}{np}\PYG{o}{.}\PYG{n}{conj}\PYG{p}{(}\PYG{n}{fftArray}\PYG{p}{)}\PYG{p}{)}\PYG{o}{.}\PYG{n}{astype}\PYG{p}{(}\PYG{n}{np}\PYG{o}{.}\PYG{n}{float}\PYG{p}{)}
\end{sphinxVerbatim}

The \sphinxcode{\sphinxupquote{mask}} deployed here is an apodized version of the input pupil.  The apodization avoids \sphinxstyleemphasis{ringing}, i.e. the appearance of Airy\sphinxhyphen{}ring\sphinxhyphen{}like structures in the power spectrum.  Apodization is achieved by modifying the input telescope pupil mask in three steps:
\begin{enumerate}
\sphinxsetlistlabels{\arabic}{enumi}{enumii}{}{.}%
\item {} 
transform it into a pure binary mask

\item {} 
apodize the binary mask by applying \sphinxhref{https://scikit-image.org/docs/dev/api/skimage.morphology.html\#skimage.morphology.binary\_dilation}{skimage.morphology.binary\_dilation} 3 times to derive 3 versions of the pupil, each 1 pixel smaller than the previous one.

\item {} 
The 1 pixel wide border region which makes the difference between successively dilated pupils is assigned a transmission value according to a Gaussian fall\sphinxhyphen{}off.

\end{enumerate}

   \begin{figure} [ht]
   \begin{center}
   \begin{tabular}{c} 
   \includegraphics[width=1.0\textwidth]{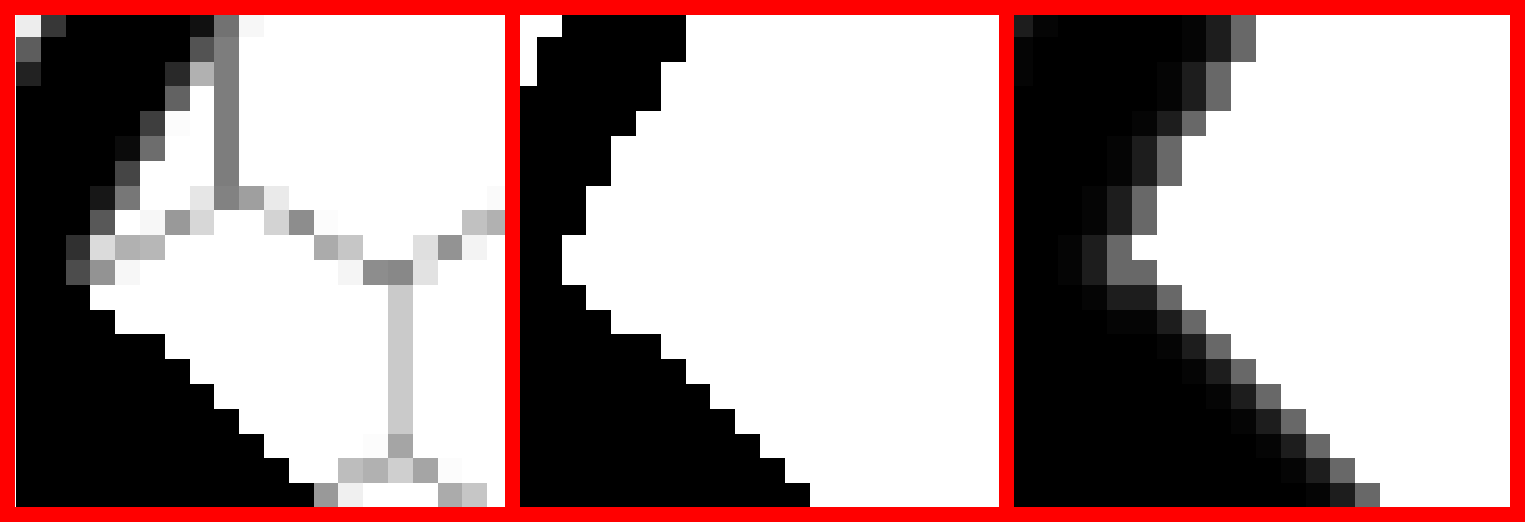}
   \end{tabular}
   \end{center}
   \caption[apodizing] 
   { \label{fig:sps:apd} 
    Pupil apodization. Left: crop of the original telescope pupil, in this case of the ELT showing a representation  of the segmentation. Center: Same crop of the pupil after “binarization”. Right: Apodized pupil applied to residual phase to avoid aliasing/ringing.}
   \end{figure}

Currently there is no parameter to vary the number of steps for the apodization. In case you absolutely want to, you’d have to assign the mask manually like so:

\begin{sphinxVerbatim}[commandchars=\\\{\}]
\PYG{g+gp}{\PYGZgt{}\PYGZgt{}\PYGZgt{} }\PYG{k+kn}{import} \PYG{n+nn}{aosat}
\PYG{g+gp}{\PYGZgt{}\PYGZgt{}\PYGZgt{} }\PYG{n}{nsteps} \PYG{o}{=} \PYG{l+m+mi}{6} \PYG{c+c1}{\PYGZsh{} let\PYGZsq{}s say you want 6 steps for a really soft pupil}
\PYG{g+gp}{\PYGZgt{}\PYGZgt{}\PYGZgt{} }\PYG{n}{a}\PYG{o}{=}\PYG{n}{aosat}\PYG{o}{.}\PYG{n}{analyze}\PYG{o}{.}\PYG{n}{sps\PYGZus{}analyzer}\PYG{p}{(}\PYG{p}{)}
\PYG{g+gp}{\PYGZgt{}\PYGZgt{}\PYGZgt{} }\PYG{n}{a}\PYG{o}{.}\PYG{n}{mask} \PYG{o}{=} \PYG{n}{util}\PYG{o}{.}\PYG{n}{apodize\PYGZus{}mask}\PYG{p}{(}\PYG{n}{a}\PYG{o}{.}\PYG{n}{sd}\PYG{p}{[}\PYG{l+s+s1}{\PYGZsq{}}\PYG{l+s+s1}{tel\PYGZus{}mirror}\PYG{l+s+s1}{\PYGZsq{}}\PYG{p}{]} \PYG{o}{!=} \PYG{l+m+mi}{0}\PYG{p}{,}\PYG{n}{steps}\PYG{o}{=}\PYG{n}{nsteps}\PYG{p}{)}
\end{sphinxVerbatim}

The resulting power spectra are averaged azimuthally, and finally temporarily.

\subsubsection{Plot captions}

   \begin{figure} [ht]
   \begin{center}
   \begin{tabular}{c} 
   \includegraphics[width=0.5\textwidth]{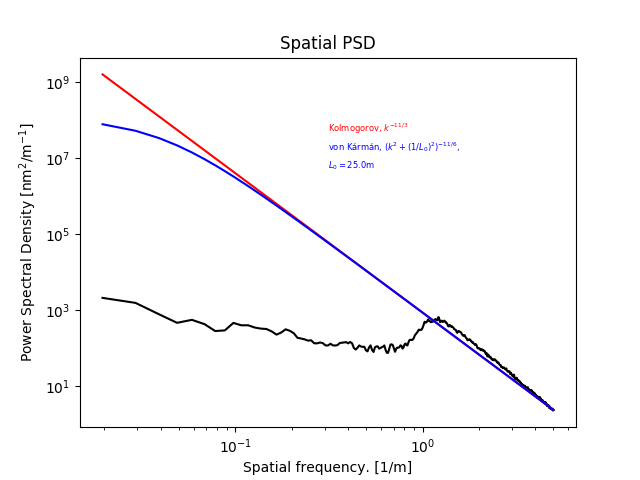}
   \end{tabular}
   \end{center}
   \caption[apodizing] 
   { \label{fig:sps} 
    Plot produced by the spatial power spectrum analyzer}
   \end{figure}

\label{\detokenize{analyzers/sps_analyzer:plot-captions}}
When called on its own mode, or on a figure with sufficient available subplot space, \sphinxcrossref{\sphinxcode{\sphinxupquote{make\_plot()}}} will produce a figure as shown in Fig.~\ref{fig:sps}. The figure caption would be:

\sphinxstyleemphasis{time\sphinxhyphen{}averaged spatial power spectrum of the residual wavefronts.  For comparison, the red line shows the expected open\sphinxhyphen{}loop Kolmogorov spectrum. The blue line represents a von Kármán spectrum for an outer scale of} \(L_0=25\) m.

Note that the blue line is plotted only, when the \sphinxcode{\sphinxupquote{L0}} key is present in the \sphinxhref{../general\_concept/setup}{setup} dictionary

\subsubsection{Resulting properties}
\label{\detokenize{analyzers/sps_analyzer:resulting-properties}}
\sphinxtitleref{sps\_analyzer} exposes the following properties after \sphinxcrossref{\sphinxcode{\sphinxupquote{finalize()}}} has been called:

\begin{savenotes}\sphinxattablestart
\centering
\sphinxcapstartof{table}
\sphinxthecaptionisattop
\sphinxcaption{sps\_analyzer properties}\label{\detokenize{analyzers/sps_analyzer:id2}}
\sphinxaftertopcaption
\begin{tabular}[t]{|\X{2}{9}|\X{2}{9}|\X{5}{9}|}
\hline
\sphinxstyletheadfamily 
Property
&\sphinxstyletheadfamily 
type
&\sphinxstyletheadfamily 
Explanation
\\
\hline
\sphinxcrossref{\sphinxcode{\sphinxupquote{mask}}}
&
2D ndarray (float)
&
Apodization mask generated from pupil
\\
\hline
\sphinxcrossref{\sphinxcode{\sphinxupquote{f\_spatial}}}
&
1D ndarray (float)
&
Spatial frequency vector {[}1/m{]}
\\
\hline
\sphinxcrossref{\sphinxcode{\sphinxupquote{ps\_psd}}}
&
1D ndarray (float)
&
Power spectral density at f\_spatial {[}nm\textasciicircum{}2/m\textasciicircum{}\sphinxhyphen{}1{]}
\\
\hline
\end{tabular}
\par
\sphinxattableend\end{savenotes}

\section{Conclusions}

We have presented AOSAT, a python package for the analysis of SCAO end-to-end simulation residual phase screens, a represnettaion of the telescope pupil and the AO parameters of the simulation. In its standard configuration, it provides a set of performance indicators and informative graphs to assess the quality of the AO correction.  We have shown how AOSAT may be used stand-alone, integrated into a simulation environment, or can easily be extended according to a user's needs. Additionally we discussed the internal workings of the included standard analyzers of AOSAT.

We hope that this package will not only produce high quality analyses for the future of our own METIS project, but will also be useful for others and perhaps even allow some comparisons between projects.

\acknowledgments 
 
AOSAT originated from a set of on-the-fly written scripts exactly as criticized in Sec.~\ref{sec:purpose}.  These scripts were used to analyze the SCAO simulation done for phase B of the METIS\cite{metis} project.

Numerous people in this project and at ESO have discussed these analyses when presented at FDR, and thus contributed to the idea of producing a coherent analysis package that will in turn produce reliable, reproducible analyses for later phases, and possibly other projects, too.

\bibliography{aosat} 
\bibliographystyle{spiebib} 

\end{document}